\documentclass[a4paper,superscriptaddress,floatfix,nofootinbib,onecolumn,longbibliography]{revtex4-1}

\usepackage{bbm}
\usepackage{bbold}
\usepackage{bbm}
\usepackage[pdftex]{graphicx}
\usepackage{latexsym,amsmath,verbatim,amssymb,txfonts}
\usepackage{lipsum}
\usepackage{color}
\usepackage{rotating}
\usepackage{verbatim}
\usepackage{multirow}
\usepackage{soul}
\usepackage[english]{babel}
\usepackage{comment}
\usepackage{bm}
\usepackage{amsmath}
\usepackage{amsfonts}
\usepackage{amssymb}
\usepackage{float}
\usepackage{color}
\usepackage{braket}
\usepackage{blkarray, bigstrut}

\newcommand\RR{\mathbb{R}}

\definecolor{internationalkleinblue}{rgb}{0.0, 0.18, 0.65}
\definecolor{applegreen}{rgb}{0.55, 0.71, 0.0}
\definecolor{darkviolet}{rgb}{0.58, 0.0, 0.83}
\definecolor{darkpink}{rgb}{0.91, 0.33, 0.5}

\begin{document}
\title{Turing patterns in systems with high-order interactions}

\author{Riccardo Muolo}
\thanks{These two authors contributed equally}
\affiliation{naXys, Namur Institute for Complex Systems, University of Namur, Belgium}
\affiliation{Department of Mathematics, University of Namur, Belgium}\email{riccardo.muolo@unamur.be}
\affiliation{Department of Applied Mathematics, Mathematical Institute, Federal University of Rio de Janeiro, Brazil}

\author{Luca Gallo}
\thanks{These two authors contributed equally}
\affiliation{naXys, Namur Institute for Complex Systems, University of Namur, Belgium}
\affiliation{Department of Physics and Astronomy \& INFN, University of Catania, Italy}

\author{Vito Latora}
\affiliation{Department of Physics and Astronomy \& INFN, University of Catania, Italy}
\affiliation{School of Mathematical Sciences, Queen Mary University of London, UK}
\affiliation{Complexity Science Hub Vienna, Austria}

\author{Mattia Frasca}
\affiliation{Department of Electrical, Electronics and Computer Science Engineering, University of Catania, Italy}
\affiliation{Istituto di Analisi dei Sistemi ed Informatica “A. Ruberti”, Consiglio Nazionale delle Ricerche (IASI-CNR), Roma, Italy}

\author{Timoteo Carletti}
\affiliation{naXys, Namur Institute for Complex Systems, University of Namur, Belgium}
\affiliation{Department of Mathematics, University of Namur, Belgium}

\date{\today}

\begin{abstract}
Turing theory of pattern formation is among the
most popular theoretical means to account for the
variety of spatio-temporal structures observed in
Nature and, for this reason, finds applications in many different
fields. While Turing patterns have been thoroughly investigated on continuous support and on networks, only a few attempts have been made towards their characterization in systems with  higher-order interactions. In this paper, we propose a way to include group  interactions in reaction-diffusion systems, and we study their effects on the formation of Turing patterns. To achieve this goal, we rewrite the problem originally studied by Turing in a general form that accounts for a microscropic description of interactions of any order in the form of a hypergraph, and we prove that the interplay between the different orders of interaction may either enhance or repress the emergence of Turing patterns. Our results shed light on the mechanisms of pattern-formation in systems with  many-body interactions and pave the way for further extensions of Turing original framework. 
\end{abstract}

\maketitle

\section{Introduction}

Many natural and engineered systems exhibit collective behaviors, manifesting themselves as spatio-temporal ordered motifs, whose emergence can be explained utterly by considering the interactions within the system \cite{anderson}. One of the most elegant and popular theories for the emergence of self-organized patterns is due to the British mathematician Alan Turing, who proposed, in the context of morphogenesis, a mechanism of pattern-formation rooted on a diffusion-driven instability, which now bears his name \cite{Turing}. In a nutshell, the Turing instability results from the combined action of two processes, (local) reaction and (long-range) diffusion, involving an activator and an inhibitor species \cite{Meinhardt,Murray2001}. While each process considered separately would drive the system to a spatially homogeneous state, there can be conditions on the models and on the interactions such that any heterogeneous, arbitrarily small, perturbation of the homogeneous state is amplified and eventually returns a macroscopic patchy (non-homogeneous) solution, i.e., a Turing pattern. The diffusive terms being the destabilizing factors, the above mechanism is also known in the literature as a diffusion-driven instability.

Despite its generality, the framework requires the involved species to diffuse with quite different rates, a condition that is not often naturally realized without introducing additional mechanisms, such as convection, electromagnetic fields, differential adherence. Indeed experimental evidence of the existence of Turing patterns was obtained almost half a century later in chemical reactions  \cite{castets,DeKepper} and in Cellular Neural Networks \cite{chua_Turing1,chua_Turing2}. Without adding unnecessary mechanisms and still looking for simple models, scholars have proposed several variants to the original Turing scheme to facilitate the emergence of patterns. For example, noise \cite{biancalani} or an upper bound to the signal propagation \cite{ZH2016,jop_carletti} can be added to the picture.

In many relevant applications by their very definition, the local reactions involve very close species, that can thus be considered spatially separated from other groups. Starting from this observation Turing idea has been extended to discrete systems: species occupy spatially limited zones, i.e., nodes, and diffuse across links connecting different zones. Initially the theory has been developed as to include regular $1D$ and $2D$ lattices \cite{OS1971}, and then extended on complex networks \cite{NM2010,pastorsatorrasvespignani2010}.

The latter framework has proved particularly fruitful for Turing patterns, especially for the introduction of asymmetric displacements \cite{Asllani2}, as in the case of directed networks \cite{Asllani1} or non-normal ones \cite{jtb}, where it was shown that Turing mechanism is enhanced. In particular, such framework is very general and allows to study the Turing instability on discrete topologies which are not trivially embedded on continuous domains, as it is the case for regular lattices. Nevertheless, certain dynamics escape the network framework, as the interactions between the elementary units are not only pairwise, but can nvolve several agents at the same time, we are thus in presence of many-body interactions. Let us here stress that we do not (only) refer to processes that introduce, a possibly small, corrections to the first order model, i.e., the pairwise coupling, but we overcome this setting and consider cases where high-order interactions are the main driver. Think of social systems where the group interaction of human being or social animals, determine the system fate in the case, e.g., of virus spreading~\cite{iacopini2019simplicial,barrat2022social}. Or consider processes where crowding is key factor; this can happen in the case of random walk~\cite{AsllaniPRL2017,AsllaniPRR2020} or chemical systems~\cite{FanelliMcKane2010,Fanelli_2013}. Let us observe that nonlinear terms introduced with the high-order interaction can be associated to nonlinear diffusion processes already studied in the literature, our results differentiate however from the latter because of the many-body assumption. One thus needs to resort to more sophisticated mathematical structures, such as hypergraphs or simplicial complexes, which are an extension of networks, beyond the framework of pairwise interactions \cite{battiston2020networks,natphys}. Dynamical systems on high-order topologies have been recently studied \cite{carletti2020dynamical,gambuzza2021stability}, but a general theory of Turing pattern formation on top of such structures is still lacking. Let us observe that a similar framework has been used to study synchronization of high-order coupled oscillators~\cite{maxime2020}. Reaction-diffusion systems have been recently studied in the frameworks of topological signals \cite{turing_topological}, which differs from the problem we are studying, as the in the former species lie in nodes and links, while in the latter only in nodes. The aim of this paper is to take one step further by proposing an extension of Turing instability on high-order topologies. For this scope, we will employ the formalism developed by Gambuzza et al. \cite{gambuzza2021stability} in the context of synchronization.

By performing a linear stability analysis involving a high-order Laplace matrix and the Jacobian matrices of the reaction and coupling parts, we will show that the joint action of high-order structures and nonlinear interactions is the key feature in driving the systems unstable. Then, considering suitable choices of couplings and high-order topologies, we have analytically determined the conditions for the instability to occur by focusing on the role of the diffusion coefficients at every order. As we will show, having different orders of diffusivity allows either to enhance or to reduce the formation of Turing patterns with respect to the case of systems with only-pairwise interactions, by tuning the parameters of the high-order couplings. 
 
The paper is organized as follows. In section \ref{sec:rdhighord} 
we will set the theoretical ground, by extending the reaction-diffusion equations originally studied by Turing to the case of systems coupled through high-order interactions. In section \ref{sec:turingth}, we will choose a particular type of nonlinear coupling, namely the diffusive-like cubic one, and we will analyze the emergence of patterns in the case of systems with only-pairwise interactions. In section \ref{sec:sezionesuHO}, we will study the effects of high-order interactions on Turing instability. We will first restrict 
to the particular case of the so-called 
natural coupling, which allows for a fully analytical treatment, showing how this is a generalization of some previous attempts to tackle the problem. We will then relax the natural coupling hypothesis, and examine an intermediate case with some specific structure, that we call regular topologies, which also allow an analytical treatment. 
Lastly, we will consider the case of the most general high-order coupling topologies, where only a numerical study is possible. Finally, in the last section, we will discuss further lines of investigation and possible applications.

\section{Reaction-Diffusion systems on high-order structures}
\label{sec:rdhighord}

Let us consider a dynamical system composed by $N$ identical units subject to some (local) nonlinear reaction dynamics. Assume moreover that 
many-body interactions, i.e., interactions in groups of more than two units, are allowed. 

Let the state of the $i$--th unit to be described by the vector $\vec{x}_i(t)\in\mathbb{R}^{m}$, then, under the mentioned assumptions, the evolution rate of the state vector of the $i$--th unit is governed by the following equation:
\begin{equation}
\frac{d{\vec{x}}_i}{dt}=\vec{f}(\vec{x_i})+\sum_{d=1}^P\sigma_d \sum_{j_1=1}^N \dots \sum_{j_d=1}^N A_{ij_1j_2\dots j_d}^{(d)}\vec{g}^{(d)}(\vec{x}_i,\vec{x}_{j_1},\vec{x}_{j_2},\dots,\vec{x}_{j_d})\, 
\label{eq:dyn1}
\end{equation}
with $i=1,\dots, N$, where $\vec{f}:\mathbb{R}^{m}\rightarrow \mathbb{R}^m$ describes the local nonlinear dynamics. Let $P+1$ denote the size of the largest group of interacting units, then $\vec{g}^{(d)}:\mathbb{R}^{m\times (d+1)}\rightarrow \mathbb{R}^m$, $d\in\{1,\dots,P\}$, are the nonlinear coupling functions ruling the $(d+1)$-body interactions, encoded into the adjacency tensors $A^{(d)}$, with $A^{(d)}_{i j_1\dots j_d}=1$ if and only if the $d+1$ units $\{i,j_1,\dots ,j_d\}$ interact together, i.e., they are connected by a hyperedge with the convention that repeated indexes yield a $0$ entry. We denote by $\sigma_d>0$ the coupling strengths. Let us assume the nonlinear couplings to be diffusive-like, hence for every $d$ there exists a function 
$\vec{h}^{(d)} \colon \RR^{m \times d} \to \RR^m$ such that $\vec{g}^{(d)}$ can be written as:
\begin{equation}
\label{eq:difflike}
\vec{g}^{(d)}(\vec{x}_i,\vec{x}_{j_1},\dots,\vec{x}_{j_d})=\vec{h}^{(d)}(\vec{x}_{j_1},\dots,\vec{x}_{j_d})-\vec{h}^{(d)}(\vec{x}_i,\dots,\vec{x}_i)\, 
\end{equation}

This definition of $\vec{g}^{(d)}$ guarantees that the coupling vanishes when all units have the same state vector, namely $\vec{g}^{(d)}(\vec{x}_i,\vec{x}_{i},\dots,\vec{x}_{i})=0$. This could be intuitively considered as a generalization of Fickean diffusion on networks, diffusion tends to homogenize local differences, hence to vanish in the case of equal system states. If, in addition, we assume the existence of $\vec{x}^*\in\mathbb{R}^{m}$ such that $\vec{f}(\vec{x}^*)=0$, meaning that a solution of the $N$ isolated systems exists; then by setting $\vec{x}_i = \vec{x}^*$, for all $i=1,\dots,N$, and by assuming Eq.~\eqref{eq:difflike} to hold true, then $(\vec{x}_1,\dots,\vec{x}_N)^\top = (\vec{x}^*,\dots,\vec{x}^*)^\top$ results to be a spatially independent solution of Eq.~\eqref{eq:dyn1}. This condition is a prerequisite for Turing instability that, as mentioned above, is based on the existence of a homogeneous stable equilibrium eventually destabilized by the diffusion. Notice that the diffusion terms are determined by the topology of the connections as well as by the parameters entering in the diffusive couplings, so that all these factors concurrently contribute to the mechanism of Turing pattern formation in the general model \eqref{eq:dyn1}.

The standard framework to study Turing instability consists of two species reaction-diffusion systems, thus even if the proposed framework is quite general, we preferred in the following to limit our analysis to a $2$-dimensional case. Let us thus set $m=2$ and denote the two components of the state vector $\vec x_i$ by $(u_i,v_i)$. Then by defining $\vec{f}(\vec{x})=(f_1(u,v),f_2(u,v))$ and $\vec{h}^{(d)}(\vec{x}_{1},\dots, \vec{x}_{d})=({h}^{(d)}_1(u_{1},\dots,u_{d},v_{1},\dots,v_{d}),{h}^{(d)}_2(u_{1},\dots,u_{d},v_{1},\dots,v_{d}))$, for all $d=1,\dots, P$, we can rewrite Eq.~\eqref{eq:dyn1} as
\begin{equation}
\label{eq:react-diff}
 \begin{cases}
 \displaystyle 
 \frac{d{u}_i}{dt}=f_1(u_i,v_i)+\sum_{d=1}^P\sigma_d\sum_{j_1=1}^N \dots \sum_{j_d=1}^N A_{i,j_1,\dots,j_d}^{(d)}\bigg[h_1^{(d)}(u_{j_1},\dots,u_{j_d},v_{j_1},\dots,v_{j_d}) \\  \qquad\qquad\qquad\qquad\qquad\qquad\qquad\qquad\qquad\qquad\qquad 
 -h_1^{(d)}(u_i,\dots,u_i,v_i,\dots,v_i)\bigg] 
 \\  \displaystyle\frac{d{v}_i}{dt}=f_2(u_i,v_i) + \sum_{d=1}^P\sigma_d\sum_{j_1=1}^N \dots \sum_{j_d=1}^N A_{i,j_1,\dots,j_d}^{(d)}\bigg[h_2^{(d)}(u_{j_1},\dots,u_{j_d},v_{j_1},\dots,v_{j_d}) 
 \\  \qquad\qquad\qquad\qquad\qquad\qquad\qquad\qquad\qquad\qquad\qquad 
 -h_2^{(d)}(u_i,\dots,u_i,v_i,\dots,v_i)\bigg]
 \end{cases}
\end{equation}
\noindent where we have taken into account condition~\eqref{eq:difflike}. To focus on the role of high-order interactions, we can further simplify the model by assuming that the nonlinear diffusion does not contain any cross-diffusion term, namely for all $d\in\{1,\dots,P\}$ the function $h_1^{(d)}$ (resp. $h_2^{(d)}$) depends only on $\{u_{j_1},\dots,u_{j_d}\}$ (resp. $\{v_{j_1},\dots,v_{j_d}\}$). A throughout analysis of the general case goes beyond the scope of this work and it could be consider in a forthcoming study. Under this hypothesis, the examined system is for all $i=1,\dots,N$
\begin{equation}
\label{eq:react-diff2}
 \begin{cases} 
 \displaystyle \frac{d{u}_i}{dt}=f_1(u_i,v_i)+\sum_{d=1}^P\sigma_d\sum_{j_1=1}^N \dots \sum_{j_d=1}^N A_{i,j_1,\dots,j_d}^{(d)}\left[ h_1^{(d)}(u_{j_1},\dots,u_{j_d}) -h_1^{(d)}(u_{i},\dots,u_{i})\right]\\ \\ \displaystyle \frac{d{v}_i}{dt}=f_2(u_i,v_i) + \sum_{d=1}^P\sigma_d\sum_{j_1=1}^N \dots \sum_{j_d=1}^N A_{i,j_1,\dots,j_d}^{(d)}\left[h_2^{(d)}(v_{j_1},\dots,v_{j_d}) -h_2^{(d)}(v_{i},\dots,v_{i}) \right] \, 
 \end{cases}
\end{equation}

In the next sections, we will study the conditions for the emergence of Turing patterns in systems of the form~\eqref{eq:react-diff2} focusing,  one-by-one, on the two novel aspects of our model. First, we will deal with the study of Turing patterns in the standard case when only two-body interactions are present (i.e., the model with $P=1$), but the diffusive 
coupling is nonlinear (Sec. \ref{sec:turingth}). Then, we  will investigate the model in presence of multi-body nonlinear diffusive coupling (Sec. \ref{sec:sezionesuHO}).

\section{Turing theory with nonlinear diffusive-like coupling}
\label{sec:turingth}

Let us start with the analysis of the conditions on the fixed point for the isolated system, that we here indicate as $(u^*,v^*)$. This means to only consider local reaction and silence the interactions among the different units. The fixed point satisfies the equations: 
\begin{equation*}
 f_1(u^*,v^*)=f_2(u^*,v^*)=0\, 
\end{equation*}
and should be stable, a condition that can be obtained by imposing
\begin{equation}\label{eq:hom_stab}\mathrm{tr}~\mathbf{J}_0<0 \text{ and }\det \mathbf{J}_0>0 \, 
\end{equation}
where $
\mathbf{J}_0=\left(\begin{matrix} \partial_u f_{1} & \partial_v f_{1} \\ \partial_u f_{2} & \partial_v f_{2} \end{matrix} \right)$ is the Jacobian matrix of the reaction function, being $\partial_a f_{\ell}=\frac{\partial f_\ell}{\partial a}(u^*,v^*)$, with $\ell\in\{1,2\}$ and $a\in\{u,v\}$. We remark that with the notation $\partial_a f_{\ell}=\frac{\partial f_\ell}{\partial a}(u^*,v^*)$ we indicate that all the derivatives in the Jacobian matrix are evaluated at the equilibrium point $(u^*,v^*)$.

Before studying the effects of high-order terms on the emergence of Turing patterns in the most general system in Eq.~\eqref{eq:react-diff2}, let us first analyze the effect of a nonlinear diffusive coupling in a system with pairwise interactions only, i.e., when $P=1$. This corresponds thus to consider a reactive system where nonlinear diffusion is present and interactions occur among units that are mapped as the nodes of a complex network. Let us observe that the problem of Turing instability with nonlinear diffusion has already been studied, but on a continuous support for the dynamics 
\cite{nonlin_Tur,nonlin_Tur2,nonlin_Tur3}. The main purpose of the following sections is to extend this analysis to the case of networked systems and to introduce the reader to the framework of Turing theory on networks.

For sake of definitiveness, we will present our results by using a cubic diffusion term. We adopt the same 
assumption even in the case of high-order interactions, so that it will be easier to examine the effects of the latter on the dynamical behavior of the system.

\subsection{Turing patterns in networked systems with nonlinear diffusion}
\label{ssec:TPnetnonlindiff}
By setting $P=1$ in Eq.~\eqref{eq:react-diff2} we obtain
\begin{equation}
\label{eq:reactnonlindiff}
\begin{cases} \displaystyle \frac{d{u}_i}{dt}=f_1(u_i,v_i)+\sigma_1\sum_{j=1}^N A_{ij}^{(1)}\left(h^{(1)}_1(u_{j})-h^{(1)}_1(u_i)\right)\\ \displaystyle \frac{d{v}_i}{dt}=f_2(u_i,v_i)+\sigma_1\sum_{j=1}^N A_{ij}^{(1)}\left(h^{(1)}_2(v_{j})-h^{(1)}_2(v_i)\right)\end{cases}\forall i=1,\dots,N\, .
\end{equation}
These equations can be linearized around the the equilibrium point $(u^*,v^*)$, by giving: 
\begin{equation}
\label{eq:linreactnonlindiff}
\begin{cases} 
\displaystyle \frac{d{\delta u}_i}{dt}=\partial_u f_1(u^*,v^*)\delta u_i+\partial_v f_1(u^*,v^*)\delta v_i+\sigma_1\sum_{j=1}^N A_{ij}^{(1)}\partial_u h^{(1)}_1(u^*)(\delta u_j-\delta u_i)\\ 
\displaystyle \frac{d{\delta v}_i}{dt}=\partial_u f_2(u^*,v^*)\delta u_i+\partial_v f_2(u^*,v^*)\delta v_i+\sigma_1\sum_{j=1}^N A_{ij}^{(1)}\partial_v h^{(1)}_2(v^*)(\delta v_j-\delta v_i)\, \end{cases}
\end{equation}
where $\delta u_i=u_i-u^*$, $\delta v_i=v_i-v^*$ and $i=1,2,\dots,N$. 
Let $L^{(1)}_{ij}=A_{ij}^{(1)}-k^{(1)}_i \delta_{ij}$ be the $i,j$ 
element of the network Laplacian matrix\footnote{Let us observe that such matrix has a non positive spectrum and it is the discrete analogous of the continuous diffusion operator $\nabla^2$ once the underlying network is a regular lattice; for this reason, the network Laplacian matrix plays a relevant role in the context of Turing pattern formation. However, we would like to point out that in the literature, e.g., consensus and synchronization, it is common to find the Laplacian matrix defined as $L^{(1)}_{ij}=k^{(1)}_i \delta_{ij}-A_{ij}^{(1)}$ (e.g., \cite{newmanbook,gambuzza2021stability}), having thus a non negative spectrum. Then in such cases the coupling term will exhibit a negative sign.}, where $k^{(1)}_i=\sum_jA_{ij}^{(1)}$ is the node degree. We can rewrite the latter equation in compact form to emphasize the $2$-dimensional nature of the problem as
\begin{equation}
\label{eq:linreactnonlindiff2}
\frac{d}{dt}\binom{\delta u_i}{\delta v_i}=\mathbf{J}_0\binom{\delta u_i}{\delta v_i}+\sigma_1 \sum_{j=1}^NL_{ij}^{(1)}\mathbf{J}_{H^{(1)}}\binom{\delta u_j}{\delta v_j}\quad\forall i=1,\dots,N\, 
\end{equation}
where $\mathbf{J}_{H^{(1)}}=\left(
\begin{smallmatrix}
 \partial_u h^{(1)}_1(u^*) & 0\\
 0 & \partial_v h^{(1)}_2(v^*)
\end{smallmatrix}
\right)$. Let us observe that matrix $\mathbf{J}_{H^{(1)}}$ is diagonal, as there are no off-diagonal terms, due to the assumption of no cross-diffusion terms.
Let $\vec{\xi}=(\delta u_1,\delta v_1,\dots,\delta u_N,\delta v_N)^\top$  we can eventually rewrite the last equation in a compact form
\begin{equation}
\label{eq:linreactnonlindiffcompact}
\frac{d\vec{\xi}}{dt}=\left(\mathbb{I}_N\otimes \mathbf{J}_0 +\sigma_1\mathbf{L}^{(1)}\otimes \mathbf{J}_{H^{(1)}} \right) \vec{\xi}\, 
\end{equation}
where $\mathbb{I}_N$ is the $N\times N$ identity matrix and $\otimes$ is the Kronecker product.

The eigenvalues of the $2N\times 2N$ linear system~\eqref{eq:linreactnonlindiffcompact} determine the stability of the solution $\vec{\xi^*}=0$, which corresponds to $(u_i,v_i)=(u^*,v^*)$ for all $i$. Being $\mathbf{L}^{(1)}$ a symmetric matrix, one can find a set of orthonormal eigenvectors $\phi_{\alpha}^{(1)}$ associated to eigenvalues $\Lambda^{(1)}_{\alpha}$, $\alpha=1,\dots,N$. We can then make one step further by projecting~\eqref{eq:linreactnonlindiffcompact} onto this basis, and thus obtaining $N$ linear and decoupled $2\times 2$ systems, each one depending on a single eigenvalue, namely
\begin{equation}
 \frac{d}{dt}\binom{\delta\hat{ u}_\alpha}{\delta\hat{ v}_\alpha}=\left[\mathbf{J}_0+\sigma_1\Lambda^{(1)}_{\alpha}\mathbf{J}_{H^{(1)}}\right] \binom{\delta\hat{ u}_\alpha}{\delta\hat{ v}_\alpha}:=\mathbf{J}^{(\alpha)}\binom{\delta\hat{ u}_\alpha}{\delta\hat{ v}_\alpha}\quad\forall \alpha=1,\dots,N\, 
\label{eq:lineqP1}
\end{equation}
Here $\delta \hat{u}_\alpha=\sum_i \delta u_i \phi^{(1)}_{\alpha,i}$ and  $\delta \hat{ v}_\alpha=\sum_i \delta v_i \phi^{(1)}_{\alpha,i}$, are the projection respectively of $\delta u_i$ and $\delta v_i$ on such eigenbasis, and the last equality defines the matrix $\mathbf{J}^{(\alpha)}$. Its eigenvalues can be obtained by solving 
\begin{equation*}
 \det \left(\mathbf{J}^{(\alpha)}-\lambda \mathbb{I}_2\right)=0\, 
\end{equation*}
that is
\begin{equation*}
\lambda^2- 2\mathrm{tr}~\mathbf{J}^{(\alpha)}\lambda+\det \mathbf{J}^{(\alpha)}=0\, 
\end{equation*}
The root with the largest real part, considered as a function of $\Lambda^{(1)}_{\alpha}$, is named dispersion relation, $\lambda_\alpha=\max\lambda\left(\Lambda^{(1)}_{\alpha}\right)$. If there exists $\hat{\alpha}$ such that $\lambda_{\hat{\alpha}}>0$, then the equilibrium solution $(u_i,v_i)=(u^*,v^*)$ is unstable and a Turing instability is observed, the system drives away from the homogeneous equilibrium to eventually reach a new, possibly heterogeneous, solution.

Let us observe that $\mathrm{tr}~\mathbf{J}^{(\alpha)}=\mathrm{tr}~\mathbf{J}_0+\Lambda^{(1)}_{\alpha}\mathrm{tr}~\mathbf{J}_H$. Because $h_1$ and $h_2$ encode a diffusive coupling, one can safely assume that $\partial_uh_1(u^*)>0$ and $\partial_vh_2(v^*)>0$, moreover being $\Lambda^{(1)}_{\alpha}\leq 0$ and $\mathrm{tr}\mathbf{J}_0<0$, one can conclude that $\mathrm{tr}\mathbf{J}^{(\alpha)}<0$ for all $\alpha$. In conclusion, a sufficient condition to have $\lambda_{\hat{\alpha}}>0$ results to be \begin{equation}\label{eq:inhom_inst}\det \mathbf{J}^{(\hat{\alpha})}<0
 \end{equation} for some $\hat{\alpha}>1$. 

\subsection{The Brusselator model with cubic diffusion}
\label{ssec:Bxlcub}

For the sake of simplicity and without loss of generality, let us illustrate the above results using as a reaction part 
the Brusselator model \cite{PrigogineNicolis1967,galla}, 
an extensively adopted dynamical system, when it comes to study the emergence of self-organized patterns. In addition, let us consider a cubic diffusion term. This accounts to set
\begin{eqnarray}
 \label{eq:brussA}
f_1(u,v)&=&1-(b+1)u+cu^2v  \nonumber\\ 
f_2(u,v) &=&bu-cu^2v \,
\end{eqnarray}
\noindent and
\begin{eqnarray}
 \label{eq:brussB}
h^{(1)}_1(u)&=&D_u^{(1)} u^3 \nonumber \\
h^{(1)}_2(u)&=&D_v^{(1)}v^3\, 
\end{eqnarray}
 where $b$ and $c>0$ are model parameters, and $D_u^{(1)}>0$ and $D_v^{(1)}>0$ are generalized diffusion coefficients. Hence, the system under investigation reads: 
\begin{equation}
\label{eq:bruss_cubic}
 \begin{cases} 
 \displaystyle \frac{d{u}_i}{dt}=1-(b+1)u_i+cu_i^2v_i+\sigma_1 D_u^{(1)}\sum_{j_1=1}^N A_{ij_1}^{(1)}(u_{j_1}^3-u_i^3) \\ \\ 
 \displaystyle \frac{d{v}_i}{dt}=bu_i-cu_i^2v_i+\sigma_1 D_v^{(1)}\sum_{j_1=1}^N A_{ij_1}^{(1)}(v_{j_1}^3-v_i^3) \, 
 \end{cases}
\end{equation}
for which it is straightforward to show the existence of a unique fixed point  $(u^*,v^*)=(1,b/c)$. Correspondingly, the matrices $\mathbf{J}_0$ and $\mathbf{J}_{H^{(1)}}$ are:  
\begin{equation}
\mathbf{J}_0=\begin{bmatrix} b-1 & c \\ -b & -c \end{bmatrix}\quad \text{and}\quad\mathbf{J}_{H^{(1)}}=3\begin{bmatrix}(u^{*})^2 D_u^{(1)} & 0 \\ 0 & (v^{*})^2 D_v^{(1)}  \end{bmatrix}\, 
\end{equation} 
Thus, the equilibrium $(u^*,v^*)$ is stable provided that $\mathrm{tr}\mathbf{J}_0=b-c-1<0$ and $\det\mathbf{J}_0=c>0$.
The conditions \eqref{eq:inhom_inst} for the onset of the instability are given by:
\begin{equation}  
\label{eq:bruss_conditions} 
\begin{cases} 
-c^3 D_u^{(1)}+(b-1){b}^2 D_v^{(1)}>0\\
 4b^2c^3 D_u^{(1)}D_v^{(1)}-{\left(-c^3 D_u^{(1)}+(b-1)b^2D_v^{(1)}\right)^{2}}<0\, 
 \end{cases} 
\end{equation} 
Let us observe that the above conditions do not depend on $\sigma_1$. As in the case of Turing instability resulting from linear diffusion, the relevant parameter is the ratio between the diffusive coefficients. 

\begin{figure*}[h]
\centering
\includegraphics[width=1.05\linewidth]{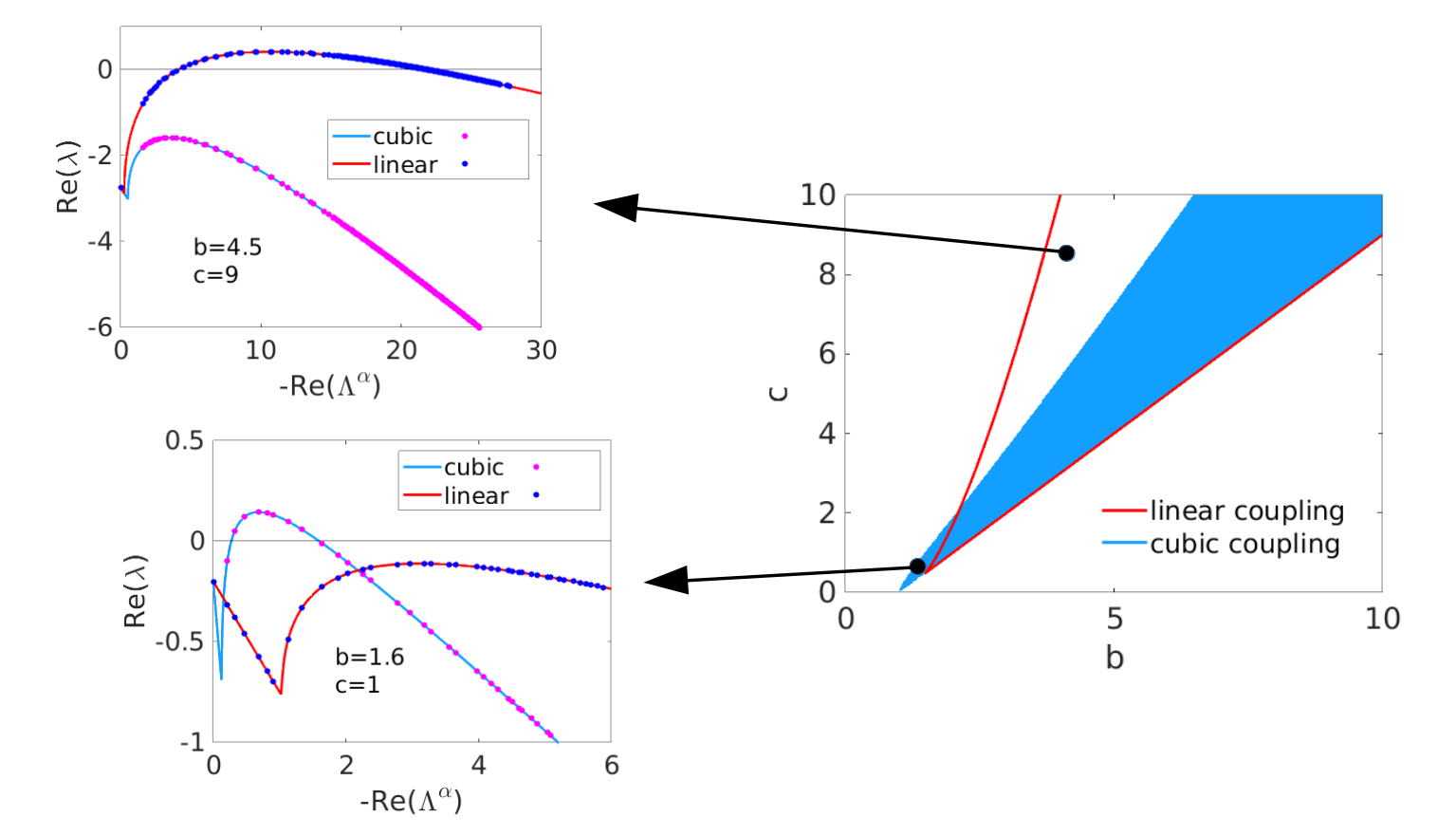}

\caption{\textit{Turing instability in networks with nonlinear diffusion.}
Right panel: Turing instability regions for the Brusselator model with $D_u^{(1)}=0.1$, $D_v^{(1)}=1$ and $\sigma_1=1$. The analytical red curves indicate the instability regions for the linear coupling, while the blue region is for the cubic coupling, as obtained numerically. We can observe that, for lower values of the parameters $(b,c)$ the cubic coupling allows for pattern formation, where the linear one does not, while for greater values we find the opposite situation. This can be visualized through the dispersion relations reported in the left panels: on the bottom left for lower values of the parameter (only the system subject to cubic coupling can go unstable), while on the upper left for greater values of the parameter (only the linear case yields patterns).
}
\label{fig:pairw_cubic}
\end{figure*}

In Fig.~\ref{fig:pairw_cubic} we compare the Turing instability regions (i.e., the set of parameters for which Turing patterns can emerge) obtained in the case of a linear diffusion term and in the case of a cubic one. By fixing in both cases the diffusion coefficients $D^{(1)}_u=0.1$ and $D^{(1)}_v=1$, and the coupling strength $\sigma_1=1$, we can observe the following behavior in the plane $(b,c)$. The instability region is wider in the case of linear coupling than in the cubic one for large values of the parameters, while the instability region shrinks to zero more slowly in the case of cubic coupling for small values of the parameters. This implies that there is a region for large enough values of $b$ and $c$ where the linear diffusion allows the emergence of Turing patterns, while in the case of cubic diffusion any initial perturbation about the homogeneous equilibrium fades out. On the contrary, if $b$ and $c$ are small enough, Turing patterns can emerge under the assumption of cubic diffusion but not if diffusion is linear. These considerations are confirmed by the dispersion relation as reported in the left panels of Fig. \ref{fig:pairw_cubic}: the reader can clearly identify the existence of eigenmodes associated to a positive dispersion relation for the cubic diffusion case (upper panel), while under the assumption of linear diffusion the dispersion relation is always negative, or the opposite case (bottom panel) where is the dispersion curve associated to nonlinear diffusion to be always negative. Let us remark that the spectrum of the Laplacian $\Lambda^{(\alpha)}$ is discrete, however in order to help the reader to better visualize the results the dispersion relations are plotted also as continuous curves.

Let us conclude this section by observing that the standard Turing framework of linear Fickean diffusion, operating both with networks or continuous supports, requires the activator species to diffuse faster than the inhibitor one~\cite{Meinhardt}, i.e., $D_v^{(1)}>D_u^{(1)}$. With the introduction of nonlinear diffusion, such condition can be relaxed, as the system can yield patterns even with equal diffusivities or with faster activator \cite{nonlin_Tur}. This claim follows directly from conditions~\eqref{eq:bruss_conditions} for the specific case of the Brusselator model. However, it is important to note that if one considers the entries of $J_{H^{(1)}}$ as the "effective" diffusion coefficients, one finds a generalization of the Turing condition $D_v^{(1)}>D_u^{(1)}$, which is \begin{equation}\label{eq:new_condition}
(v^{*})^2 D_v^{(1)}>(u^{*})^2 D_u^{(1)}
\end{equation} for the cubic case. In fact, conditions \eqref{eq:hom_stab} and \eqref{eq:inhom_inst} imply that $\partial_u f_1 < |\partial_v f_2|$, from which we obtain \begin{displaymath}
\partial_u f_1 (v^{*})^2 D_v^{(1)}  > \partial_v f_2 (u^{*})^2 D_u^{(1)} \Rightarrow \partial_u f_1 (v^{*})^2 D_v^{(1)} > |\partial_v f_2| (u^{*})^2 D_u^{(1)} > \partial_u f_1 (u^{*})^2 D_u^{(1)}
\end{displaymath} hence the new condition $(v^{*})^2 D_v^{(1)}>(u^{*})^2 D_u^{(1)}$.\\

Let us remark that the above condition is necessary but sufficient for Turing patterns. This means that if conditions \eqref{eq:bruss_conditions} are verified, then we have condition \eqref{eq:new_condition}, but the vice versa is not true.

 In the following, we will make use of the above results and focus our attention on the impact of high-order terms on the onset of the instability. The nonlinear coupling is necessary to make such many-body interactions meaningful, as pointed out in \cite{neuhauser2020multibody}: in fact, if the high-order coupling is linear, high-order interactions are nothing more than the sum of pairwise ones. Moreover, nonlinear couplings are of particular physical interest, as they are associated to anomalous diffusion \cite{sadighi2007exact}.

\section{High-order interactions}
\label{sec:sezionesuHO}

 Let us now consider the most general case of Eq.~\eqref{eq:react-diff2} with 
  high-order interactions. 
  For simplicity and without loss of generality, we will limit ourselves to consider only first and second order interactions, i.e., $P=1$ and $P=2$. Nevertheless, the theory goes beyond the exposed examples and can be straightforwardly generalized as to include higher $P$. The system we are thus interested in is given by
\begin{equation}
\label{eq:bruss_HG}
 \begin{cases} \displaystyle \frac{d{u}_i}{dt}=f_1(u_i,v_i)+\sigma_1 D_u^{(1)}\sum_{j_1=1}^N A_{ij_1}^{(1)}(h^{(1)}_1(u_{j_1})-h^{(1)}_1(u_i))\\
 \qquad\qquad\qquad\qquad\qquad\qquad +\sigma_2 D_u^{(2)}\sum\limits_{j_1=1}^N\sum\limits_{j_2=1}^N A_{ij_1 j_2}^{(2)}(h^{(2)}_1(u_{j_1},u_{j_2})-h^{(2)}_1(u_i,u_i)) \\ \displaystyle \frac{d{v}_i}{dt}=f_2(u_i,v_i)+\sigma_1 D_v^{(1)}\sum_{j_1=1}^N A_{ij_1}^{(1)}(h^{(1)}_2(v_{j_1})-h^{(1)}_2(v_i)) \\
 \qquad\qquad\qquad\qquad\qquad\qquad +\sigma_2 D_v^{(2)}\sum\limits_{j_1=1}^N\sum\limits_{j_2=1}^N A_{ij_1 j_2}^{(2)}(h^{(2)}_2(v_{j_1},v_{j_2})-h^{(2)}_2(v_i,v_i))\,  \end{cases}
\end{equation}
where $h^{(1)}_1(u)$ and $h^{(1)}_2(v)$ encode the first order interaction, while $h^{(2)}_1(u_1,u_2)$ and $h^{(2)}_2(v_1,v_2)$ model the second order coupling. 
We again assume the existence of a homogeneous solution $(u_i,v_i)=(u^*,v^*)$ and assume it to be stable once we silence both the pairwise coupling, $P=1$, and the high-order one, $P>1$. To study its stability under spatially dependent perturbations, we can follow the derivation presented in the previous section (see also~\cite{gambuzza2021stability}), that ultimately relies on the computation of a Master Stability Function~\cite{pecora1998master} in a setting involving a stationary equilibrium and high-order coupling.

Let us introduce again the perturbation vector $\vec{\xi}=(\delta u_1,\delta v_1,\dots,\delta u_N,\delta v_N)^\top$, where $\delta u_i=u_i-u^*$ and $\delta v_i=v_i-v^*$, then one can straightforwardly show that it evolves according to 
\begin{equation}
\label{eq:general_high_order_lin_eq}
\frac{d\vec{\xi}}{dt}=\left(\mathbb{I}_N\otimes \mathbf{J}_0 +\sigma_1\mathbf{L}^{(1)}\otimes \mathbf{J}_{H^{(1)}}+\sigma_2\mathbf{L}^{(2)}\otimes \mathbf{J}_{H^{(2)}} \right) \vec{\xi}\, 
\end{equation}
where $\mathbf{J}_{H^{(2)}}=\left(
\begin{smallmatrix}
 \partial_{u_1} h^{(2)}_1(u^*,u^*)+\partial_{u_2} h^{(2)}_1(u^*,u^*) & 0\\
 0 & \partial_{v_1} h^{(2)}_2(v^*,v^*)+\partial_{v_2} h^{(2)}_2(v^*,v^*)
\end{smallmatrix}\right)$ and $ \mathbf{L}^{(2)}$ represents a generalized Laplacian matrix \cite{gambuzza2021stability} accounting for the three-body interactions, whose elements are given by 
\begin{equation}
\label{eq:2-laplacian}
\mathbf{L}^{(2)}_{ij} = \begin{cases}
-\sum_{j,k=1}^{N} A^{(2)}_{ijk} & \mathrm{for} \ i=j \\ \\
\phantom{+} \sum_{k=1}^{N} A^{(2)}_{ijk} & \mathrm{for} \ i\neq j \\
\end{cases}
\end{equation}

In the following, we will examine different cases of coupling: first, in section \ref{sec:natural}, the so called natural coupling \cite{gambuzza2021stability, carletti}, which allows an analytical treatment of the problem, but does not permit to fully unveil the potential of high-order terms. The latter will be fully exploited in section \ref{sec:special}, where we will present two special cases of high-order structures in which, despite the general form of the coupling functions, a complete analytical study can be performed. Finally, in section \ref{sec:general} the scenario where both the coupling and the topology are general will be dealt with.

\subsection{Natural coupling}\label{sec:natural}
A largely used assumption on the coupling terms $\vec{h}^{(d)}$, is that once they are evaluated on the synchronous manifold, they return the same value given by $\vec{h}^{(1)}$, more precisely 
\begin{equation*}
\vec{h}^{(d)}(\vec{x},\dots,\vec{x})=...=\vec{h}^{(2)}(\vec{x},\vec{x})=\vec{h}^{(1)}(\vec{x})\, 
\end{equation*}
Such condition is known in the literature as {\em natural coupling} \cite{gambuzza2021stability}. In the present framework this implies that 
\begin{equation*}
{h}_1^{(2)}(u,u)={h}_1^{(1)}(u)\text{ and }{h}_2^{(2)}(v,v)={h}_2^{(1)}(v)\, 
\end{equation*}

To compare the case $P>1$ with the one presented for $P=1$, we consider nonlinearities based on polynomials and product of variables with total power equal to three, both in the first and in the second order terms. In particular, we set $h^{(1)}_1(u) = D^{(1)}_u u^3$ and $h^{(1)}_2(v) = D^{(1)}_v v^3$ for the two-body coupling, while we take $h^{(2)}_1(u_1,u_2)=D^{(2)}_u u_1^2u_2$ and $h^{(2)}_2(v_1,v_2)=D^{(2)}_v v_1^2v_2$, for the three-body interaction. Note that this choice for the coupling functions satisfies the condition of natural coupling as long as
\begin{equation}\label{eq:equal_coeff}
D_u^{(1)}=D_u^{(2)} \text{ and } D_v^{(1)}=D_v^{(2)} \, 
\end{equation}

Given these coupling functions, we can rewrite Eq.~\eqref{eq:bruss_HG} as
\begin{equation}
\label{eq:bruss_HG_cubic}
 \begin{cases} \displaystyle \frac{d{u}_i}{dt}=f_1(u_i,v_i)+\sigma_1 D_u^{(1)}\sum_{j_1=1}^N A_{ij_1}^{(1)}(u_{j_1}^3-u_i^3)+\sigma_2 D_u^{(2)}\sum\limits_{j_1=1}^N\sum\limits_{j_2=1}^N A_{ij_1 j_2}^{(2)}(u_{j_1}^2u_{j_2}-u_i^3) \\ \displaystyle \frac{d{v}_i}{dt}=f_2(u_i,v_i)+\sigma_1 D_v^{(1)}\sum_{j_1=1}^N A_{ij_1}^{(1)}(v_{j_1}^3-v_i^3) +\sigma_2 D_v^{(2)}\sum\limits_{j_1=1}^N\sum\limits_{j_2=1}^N A_{ij_1 j_2}^{(2)}(v_{j_1}^2v_{j_2}-v_i^3)\,  \end{cases}
\end{equation}

Again, to study the stability of the homogeneous solution $(u_i,v_i)=(u^*,v^*)$ we linearize Eq.~\eqref{eq:bruss_HG_cubic}. By resorting to the natural coupling assumption we can conclude that $\mathbf{J}_{H^{(1)}}=\mathbf{J}_{H^{(2)}}$ and thus rewrite Eq.~\eqref{eq:general_high_order_lin_eq} as follows:
\begin{equation}
\label{eq:natural_lin}
\frac{d\vec{\xi}}{dt}=\left[\mathbb{I}_N\otimes \mathbf{J}_0 +\left(\sigma_1\mathbf{L}^{(1)}+\sigma_2\mathbf{L}^{(2)}\right)\otimes \mathbf{J}_{H^{(1)}} \right] \vec{\xi}\, 
\end{equation}
where 
\begin{equation*}
\mathbf{J}_{H^{(1)}}=3\begin{bmatrix}(u^{*})^2 D_u^{(1)} & 0 \\ 0 & (v^{*})^2 D_v^{(1)}  \end{bmatrix}\, 
\end{equation*} and $\vec{\xi}=(\delta u_1,\delta v_1,\dots,\delta u_N,\delta v_N)^\top$, with $\delta u_i=u_i-u^*$ and $\delta v_i=v_i-v^*$. Let us observe that Eq.~\eqref{eq:natural_lin} recalls the similar result obtained in~\cite{carletti2020dynamical}. However, the mathematical framework here proposed is more general, as it allows to study a broader class of high-order couplings (see Appendix \ref{sec:appA} for more details). 

\begin{figure*}[h!]
\centering
\includegraphics[width=0.4\linewidth]{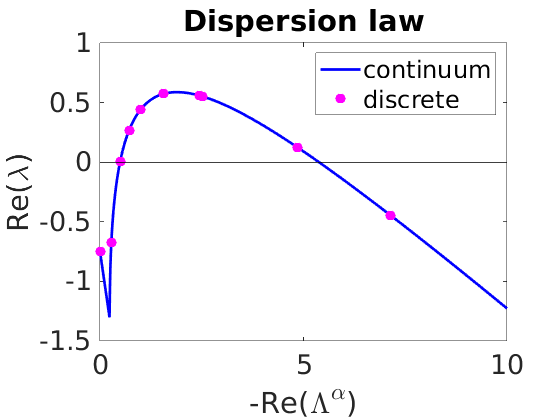}
\includegraphics[width=0.4\linewidth]{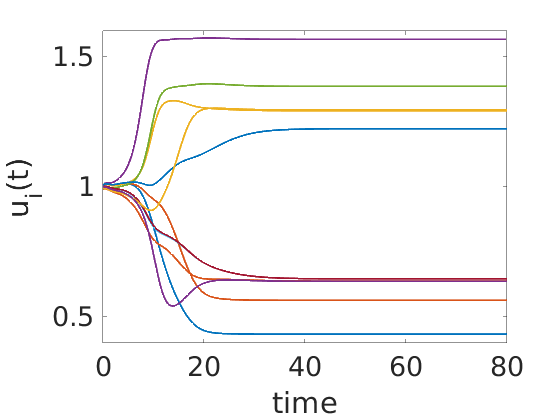}
\includegraphics[width=0.4\linewidth]{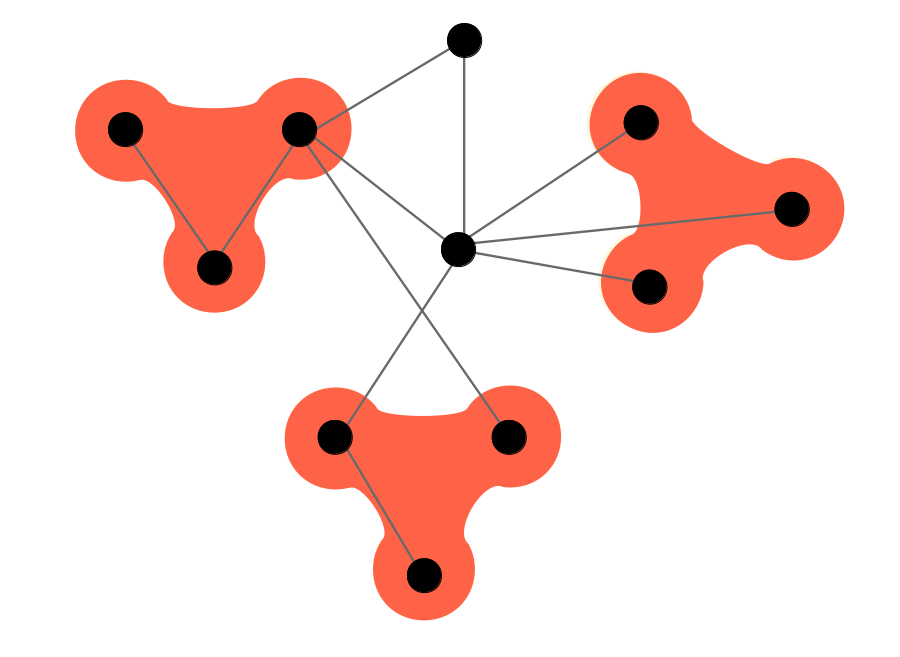}

\caption{\textit{Turing instability with nonlinear diffusion in high-order structures.}
Brusselator model with $b=3$, $c=3.5$, $D_u^{(1,2)}=0.1$, $D_v^{(1,2)}=2$, $\sigma_{1,2}=1$; the initial perturbation is $\sim 10^{-2}$. On the upper left panel, the dispersion law: the blue line is the continuous one, while the cyan dots indicate the discrete counterpart. On the upper right panel, Turing patterns for the $u$ species. On the bottom panel the hypergraph on which the system has been simulated: $11$ nodes, $11$ links and $3$ $2$-hyperedges (i.e., triangles).}
\label{fig:nat_coupl}
\end{figure*} 

The natural coupling assumption allowed us to introduce an ``effective'' Laplacian that encodes the high-order structure into a weighted network, whose weights are self-consistently determined, $\mathbf{M}=\sigma_1\mathbf{L}^{(1)}+\sigma_2\mathbf{L}^{(2)}$. Resorting to the eigenbasis for $\mathbf{M}$ one can make one step further and project the $2N\times 2N$ linear system onto $N$ linear systems of size $2\times 2$ depending each one on a single eigenvalue, $\Lambda_{\alpha}$, of $\mathbf{M}$, more precisely
\begin{equation}\label{eq:mse_natcoupl}
 \frac{d}{dt}\binom{\delta \hat{ u}_\alpha}{\delta \hat{ v}_\alpha}=\left[\mathbf{J}_0+\Lambda_{\alpha}\mathbf{J}_{H^{(1)}}\right] \binom{\delta \hat{u}_\alpha}{\delta \hat{ v}_\alpha}:=\mathbf{J}^{(\alpha)}\binom{\delta \hat{ u}_\alpha}{\delta \hat{v}_\alpha}\quad\forall \alpha=1,\dots,N\, 
\end{equation}
where again $\delta \hat{ u}_\alpha=\sum_i \delta u_i \phi_{\alpha,i}$, resp. $\delta \hat{ v}_\alpha=\sum_i \delta v_i \phi_{\alpha,i}$, is the projection of $\delta u_i$, resp. $\delta v_i$, on the eigenvector $\phi_{\alpha}$. 

Let us observe that the latter equation is formally the same as the one obtained in the case of pairwise interactions~\eqref{eq:lineqP1} and thus the same analysis follows, except that now the eigenvalues depend on the coupling strengths. In Fig.~\ref{fig:nat_coupl}, we show the dispersion relation and an example of Turing patterns for the Brusselator model (Eqs. \eqref{eq:brussA}). Let us notice that the blue curve (left panel Fig.~\ref{fig:nat_coupl}) has been computed by considering $\Lambda_{\alpha}$ to be a continuous variable, corresponding thus to the dispersion relation for a system defined on a continuous support with periodic boundary conditions. Having fixed the topology of binary interactions and the high-order ones, hence the matrices $\mathbf{L}^{(1)}$ and $\mathbf{L}^{(2)}$, we can vary the coupling strengths $\sigma_{1}$ and $\sigma_{2}$, and thus letting $\Lambda_{\alpha}$ ``to slide'' along this curve (cyan dots). Stated differently, the Turing instability can be obtained (or repressed) by simply changing the coupling and keeping the remaining part of the model unchanged.

\subsection{Regular topologies} \label{sec:special}

Let us now relax the natural coupling assumption by setting different diffusion coefficients at the different considered  order, i.e., $D_u^{(1)}\neq D_u^{(2)}$ and $D_v^{(1)}\neq D_v^{(2)}$, but keeping the same form for the coupling functions.  Namely we take $h^{(1)}_1(u) = D^{(1)}_u u^3$ and $h^{(1)}_2(v) = D^{(1)}_v v^3$ for the two-body coupling, while we take $h^{(2)}_1(u_1,u_2)=D^{(2)}_u u_1^2u_2$ and $h^{(2)}_2(v_1,v_2)=D^{(2)}_v v_1^2v_2$ for the three-body coupling. It would also be  possible to adopt different coupling functional forms for each order, but, as it will be clear in the following, it is more interesting to focus on the diffusion coefficients, the latter having a key role in the Turing mechanism of pattern formation. Nevertheless, the following analysis could be easily extended to a setting in which coupling functions have different form, as long as they remain diffusive-like. 

The starting point for the linear stability analysis is again Eq.~\eqref{eq:general_high_order_lin_eq}; however, because $\mathbf{J}_{H^{(1)}}\neq \mathbf{J}_{H^{(2)}}$ one cannot simplify the latter equation to obtain the analogous of~\eqref{eq:natural_lin}. In the same spirit of simplifying the stability equation, we can observe that in general the two Laplacians do not commute and thus they cannot be simultaneously diagonalized. In conclusion one cannot determine a single equation, depending on the spectrum of the involved Laplacian matrices, and proceed with the analysis as done in the case of the Master Stability Function. The numerical computation of the eigenvalues of the $2N\times 2N$ linear system would not allow a clear understanding of the role of the involved parameters.

There are however some high-order structures allowing for a complete analytical description: this is the case whenever the high-order Laplacian  matrix is a multiple of the network Laplacian one. We will call topologies with the above property \textit{regular topologies}. Tetrahedra or icosaedra are examples of regular topologies as well as the \textit{triangular lattice with periodic boundary conditions}, i.e., a $2$-torus paved with triangles, that we hereby analyze in details. Let us observe that we assume the nodes forming a triangle (of first order interactions) to be also part of a three-body interaction; for this reason we will call it triangular $2$-lattice. In this case, each node interacts with its six neighbors through six two-body interactions and six three-body interactions

Let us consider a triangular $2$-lattice of $N$ nodes, where each node $i$ has $6$ incident links. Its first order Laplacian will then be \begin{displaymath}
\textbf{L}_{ij}^{(1)}=\begin{cases} -6 ~~\mbox{ if }~ i=j \\ A_{ij}^{(1)}~~\mbox{ otherwise}\end{cases}
\end{displaymath}

If now each node will also have $6$ incident triangles, we have that \begin{displaymath} \sum_{j,k=1}^{N} A^{(2)}_{ijk}= 2\cdot 6 =12~~~\mbox{ and }~~~  \sum_{k=1}^{N} A^{(2)}_{ijk} = 2A^{(1)}_{ij} \end{displaymath} 
From the definition of the second order Laplacian, Eq. \eqref{eq:2-laplacian}, we obtain the relation

\begin{equation}
\label{eq:2-laplacian_1-laplacian}
\mathbf{L}^{(2)}=2 \mathbf{L}^{(1)}
\end{equation}
hence the two Laplacians can be both diagonalized through the eigenvectors of $\mathbf{L}^{(1)}$. \\
 Eq.~\eqref{eq:general_high_order_lin_eq} now takes the form
\begin{equation}
\label{eq:natural_lin_tri}
\frac{d\vec{\xi}}{dt}=\left(\mathbb{I}_N\otimes \mathbf{J}_0 + \mathbf{L}^{(1)} \otimes (\sigma_1 \mathbf{J}_{H^{(1)}}+2\sigma_2 \mathbf{J}_{H^{(2)}})\right) \vec{\xi}\, 
\end{equation}
We can then proceed in projecting on the eigenvectors of $\mathbf{L}^{(1)}$, by obtaining
\begin{equation}\label{eq:mse_trig}
 \frac{d}{dt}\binom{\delta \hat{ u}_\alpha}{\delta \hat{ v}_\alpha}=\left[\mathbf{J}_0+\Lambda_{\alpha}(\sigma_1 \mathbf{J}_{H^{(1)}}+2\sigma_2 \mathbf{J}_{H^{(2)}})\right] \binom{\delta \hat{ u}_\alpha}{\delta \hat{ v}_\alpha}\quad\forall \alpha=1,\dots,N\, 
\end{equation}
which is analogous to Eq. \eqref{eq:mse_natcoupl} due to the properties of the topology. The Laplacian eigenvalues $\Lambda_{\alpha}$ can be effectively replaced by a continuous parameter, which is the continuous spectrum of the corresponding diffusion operator, allowing us to have an analytical expression of the dispersion law. The above result can be straightforwardly generalized to all regular topologies..

\begin{figure*}[h!]
\centering
\includegraphics[width=0.4\linewidth]{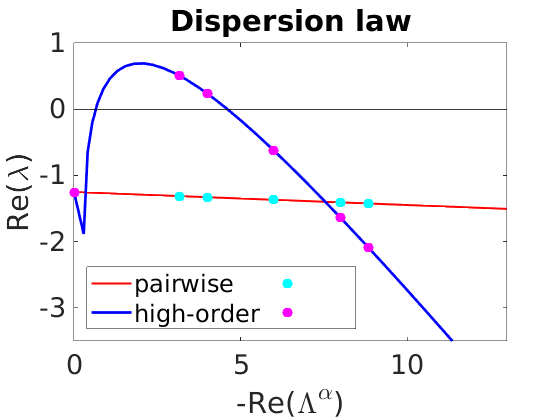}
\includegraphics[width=0.4\linewidth]{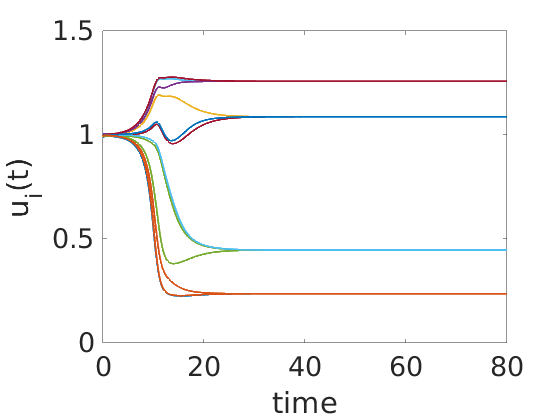}
\caption{Triangular $2$-lattice of $16$ nodes with periodic boundary conditions; Brusselator model with $b=5.5$, $c=7$, $D_u^{(1)}=1$, $D_v^{(1)}=0.5$, $D_u^{(2)}=0.1$, $D_v^{(2)}=1$, $\sigma_1=0.01$ and $\sigma_2=1$; the initial perturbation is $\sim 10^{-2}$. In the left panel, the dispersion law for the high-order case (blue line and magenta dots) compared with the case where only pairwise interactions are present (purple dashed line). In the right panel, an example of a Turing pattern for the $u$ species.}
\label{fig:nonnat_coupl_trig}
\end{figure*}

Let us point out that the assumption of $\mathbf{L}^{(2)}$ to be proportional to $\mathbf{L}^{(1)}$ allows to emphasize the effects of high-order interactions which would not be possible to fully appreciate under the natural coupling assumption. In fact, one can deal with sets of parameters which do not allow the formation of patterns by considering the sole pairwise interactions, i.e., condition \eqref{eq:new_condition} would not be satisfied, $(v^{*})^2 D_v^{(1)}<(u^{*})^2 D_u^{(1)}$, and thus patterns could not develop, but the appropriate values of the second order couplings would make possible the Turing instability. This is the case shown in Fig. \ref{fig:nonnat_coupl_trig}, where we can observe the dispersion law and an example of a pattern obtained with the Brusselator model on a triangular $2$-lattice of $16$ nodes; we notice that, while the pairwise case (purple dashed line) never allows instability, the high-order one yields Turing patterns. Let us remark that the dispersion law is a function of the spectrum of the Laplacian $\textbf{L}^{(1)}$.

\begin{figure*}[h!]
\centering
\includegraphics[width=0.4\linewidth]{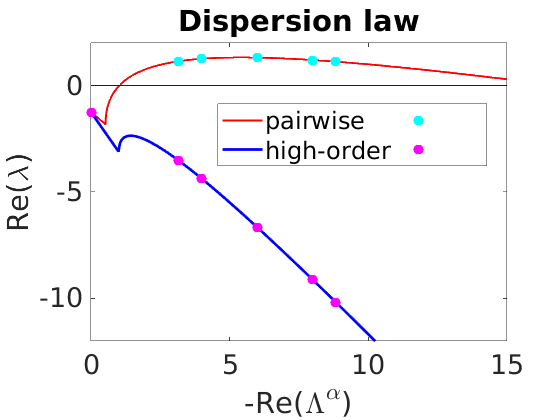}
\includegraphics[width=0.4\linewidth]{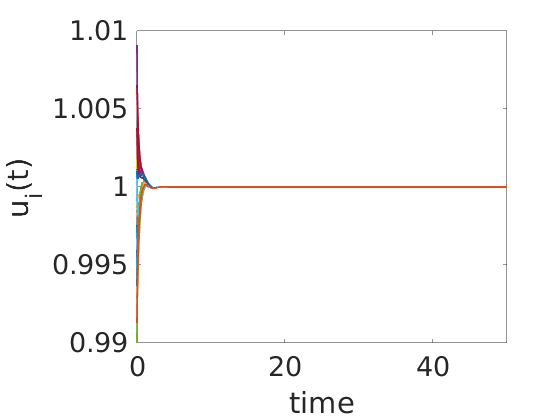}
\caption{Triangular $2$-lattice of $16$ nodes with periodic boundary conditions; Brusselator model with $b=5.5$, $c=7$, $D_u^{(1)}=0.1$, $D_v^{(1)}=1.5$, $D_u^{(2)}=1$, $D_v^{(2)}=0.5$, $\sigma_1=0.7$ and $\sigma_2=0.2$; the initial perturbation is $\sim 10^{-2}$. In the left panel, the dispersion law for the high-order case (blue line and magenta dots) compared with the case where only pairwise interactions are present (purple dashed line). As the high-order dispersion law is always negative, there is no emergence of Turing patterns, as also shown in the right panel for the $u$ species.}
\label{fig:nonnat_coupl_trig2}
\end{figure*}

Let us stress that it is also possible to find the opposite case, in which pairwise interactions would normally give rise to Turing patterns, but the presence of high-order ones annihilates them and stabilizes the system, as exemplified in Fig. \ref{fig:nonnat_coupl_trig2}. \\

The interplay between nonlinear diffusion and regular topologies could be better highlighted by rewriting \eqref{eq:mse_trig}. Under our working assumption we have
\begin{equation}
\mathbf{J}_{H^{(1)}}=3\begin{bmatrix}D_u^{(1)} & 0 \\ 0 & \Big(\frac{b}{c}\Big)^2 D_v^{(1)}  \end{bmatrix} \quad\text{ and } \quad \mathbf{J}_{H^{(2)}}=3\begin{bmatrix}D_u^{(2)} & 0 \\ 0 & \Big(\frac{b}{c}\Big)^2 D_v^{(2)}  \end{bmatrix}\, .
\end{equation}
We can hence define the effective diffusion coefficients 
\begin{equation}
\mathcal{D}_u^{eff}=\sigma_1 D_u^{(1)}+\sigma_2 D_u^{(2)}\quad\text{ and }\quad \mathcal{D}_v^{eff}=\sigma_1 D_v^{(1)}+\sigma_2 D_v^{(2)}
\end{equation}
so that we can cast~\eqref{eq:natural_lin_tri} in the following way
\begin{equation}
\label{eq:natural_lin_tri_new}
\frac{d\vec{\xi}}{dt}=\left(\mathbb{I}_N\otimes \mathbf{J}_0 + \mathbf{L}^{(1)} \otimes \mathbf{J}_{\mathcal{H}}^{eff}\right) \vec{\xi}\, ,
\end{equation} where 
\begin{equation} 
\mathbf{J}_{\mathcal{H}}^{eff} = 3\begin{bmatrix}\mathcal{D}_u^{eff} & 0 \\ 0 & \Big(\frac{b}{c}\Big)^2 \mathcal{D}_v^{eff}  
\end{bmatrix}\, .
\end{equation} 
By projecting on the eigenvectors of $\mathbf{L}^{(1)}$, we obtain a new form of Eq.~\eqref{eq:mse_trig}, namely
\begin{equation}\label{eq:mse_trig_new}
 \frac{d}{dt}\binom{\delta \hat{ u}_\alpha}{\delta \hat{ v}_\alpha}=\left[\mathbf{J}_0+\Lambda_{\alpha}\mathbf{J}_{\mathcal{H}}^{eff}\right] \binom{\delta \hat{ u}_\alpha}{\delta \hat{ v}_\alpha}\quad\forall \alpha=1,\dots,N\, .
\end{equation}
 Following the steps presented above to determine the onset of Turing instability, we can now obtain
\begin{equation}  
\label{eq:bruss_conditions_new} 
\begin{cases}
-c^3 \mathcal{D}_u^{eff}+(b-1){b}^2 \mathcal{D}_v^{eff} >0\\
 4b^2c^3 \mathcal{D}_u^{eff}\mathcal{D}_v^{eff} -{\left(-c^3 \mathcal{D}_u^{eff}+(b-1)b^2\mathcal{D}_v^{eff} \right)^{2}}<0\, .
 \end{cases} 
\end{equation} 
From the latter relations it emerges that Turing patterns are the result of the interplay not only between the model parameters and the diffusion coefficients, but also with the strength of the interactions at every order.

In the next section, we will study the general case where, in absence of specific assumptions on the coupling or on the structure, an analytical study cannot be performed, and we thus have to resort to numerical simulations. But before that, let us examine a particular case of regular topology, were the Laplacians of all orders can be diagonalized simultaneously:
the \textit{all-to-all coupling}, where every possible $d$-body interaction is active. When dealing with such coupling, the analysis is not restricted to some orders of interactions but can be extended to all of them. Nonetheless, without further ado, let us again restrict the analysis to first- and second-order interactions. For $N$ interconnected systems we have $\mathbf{L}^{(2)}=(N-2)\mathbf{L}^{(1)}$ (see \cite{gambuzza2021stability} for a detailed analysis); let us observe that this is again a case where $\mathbf{L}^{(2)}$ is proportional to $\mathbf{L}^{(1))}$, however here the proportionality constant depends on the system size. Based on the above, Eq.~\eqref{eq:general_high_order_lin_eq} can be rewritten as
\begin{equation}
\label{eq:natural_lin_a2a}
\frac{d}{dt}\vec{\xi}=\left(\mathbb{I}_N\otimes \mathbf{J}_0 + \sigma_1 \mathbf{L}^{(1)} \otimes \left(\mathbf{J}_{H^{(1)}}+\frac{\sigma_2}{\sigma_1} (N-2) \mathbf{J}_{H^{(2)}}\right)\right) \vec{\xi}\, 
\end{equation}
Projecting again on the eigenvectors of $\mathbf{L}^{(1)}$ we obtain
\begin{equation*}
 \frac{d}{dt}\binom{\hat{\delta u}_\alpha}{\hat{\delta v}_\alpha}=\left[\mathbf{J}_0+\sigma_1\Lambda_{\alpha}\left(\mathbf{J}_{H^{(1)}}+\frac{\sigma_2}{\sigma_1}(N-2) \mathbf{J}_{H^{(2)}}\right)\right] \binom{\hat{\delta u}_\alpha}{\hat{\delta v}_\alpha}\quad\forall \alpha=1,\dots,N\, 
\end{equation*}

Let us observe that $\Lambda_{\alpha}\in \{0,N\}$, hence the dispersion law depends on the number of nodes. The above equation indicates that it is possible to decompose the instability modes. 

For a given value of $N$, having fixed the ratio $\sigma_2/\sigma_1$, we can vary the value of $\sigma_{1}$, thus generating a continuous curve. Observe that such a curve allows us to check for the onset of Turing instability on the ``all-to-all'' configuration. Indeed, when the curve is negative Turing instability cannot occur for any value of $\sigma_1$, while Turing patterns can emerge in the opposite case. Hence, the continuous curve, as function of a parameter $\gamma = \sigma_1\Lambda_\alpha$, can be subsumed in the framework of the Master Stability Function approach \cite{pecora1998master}.

\begin{figure*}[h!]
\centering
\includegraphics[width=0.4\linewidth]{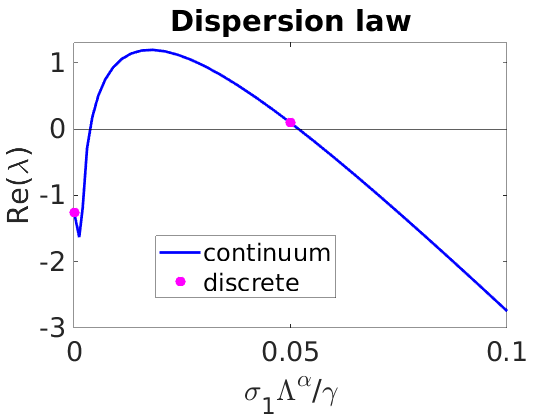}
\includegraphics[width=0.4\linewidth]{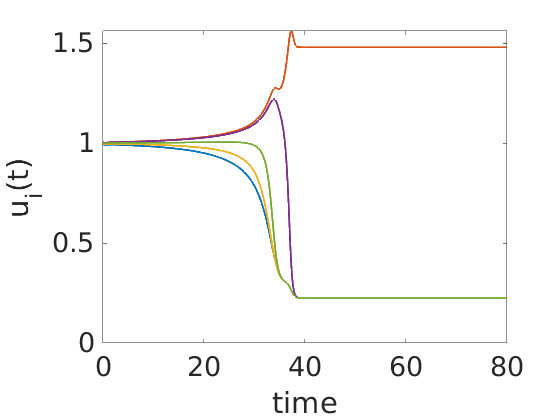}
\caption{All-to-all 2-hypergraph of $5$ nodes; Brusselator model with $b=5.5$, $c=7$, $D_u^{(1)}=1$, $D_v^{(1)}=0.1$, $D_u^{(2)}=0.07$, $D_v^{(2)}=1$, $\sigma_1=0.01$ and $\sigma_2=1$; the initial perturbation is $\sim 10^{-2}$. On the left panel, the discrete dispersion law; note that the continuous curve is a fictitious dispersion law, but it is a Master Stability Function, as explained in the text; for this reason the continuous curve can be thought as a function of a parameter $\gamma$. On the right panel, an example of a Turing pattern for the $u$ species.}
\label{fig:nonnat_coupl}
\end{figure*} 

To provide a numerical example, let us consider an ensemble of $N$ Brusselator systems interacting in the all-to-all configuration, with $h^{(1)}_1 = D^{(1)}_u u^3$, $h^{(1)}_2 = D^{(1)}_v v^3$, $h^{(2)}_1(u_1,u_2)=D^{(2)}_u u_1^2u_2$ and $h^{(2)}_2(v_1,v_2)=D^{(2)}_v v_1^2v_2$, and with the values of the diffusion coefficients set such that the coupling functions do not satisfies the natural coupling condition. In Fig.~\ref{fig:nonnat_coupl}, we report a case analogous to that displayed in Fig.~\ref{fig:nonnat_coupl_trig}, namely patterns emerge due to high-order interaction. Also in this case, with the appropriate coupling functions we can obtain the opposite situation, i.e., high-order interactions annihilate patterns, which would otherwise emerge with the sole pairwise ones.

\subsection{General topologies} \label{sec:general}

Let us now focus on the most general case, in which arbitrary high-order topologies are considered and the coupling is not restricted to the natural coupling form. Following the same reasoning of the previous section,we use different diffusion coefficients, while the coupling functions remain of the same form. 

In such setting, we cannot have a semi-analytical form of the dispersion law, because neither the Laplacians nor the matrices $J_{H^{(d)}}$ can be diagonalized simultaneously. Hence, we will resort to numerical simulations to determine the onset of the Turing instability.\\
\begin{figure*}[h!]
\centering
\includegraphics[width=0.45\linewidth]{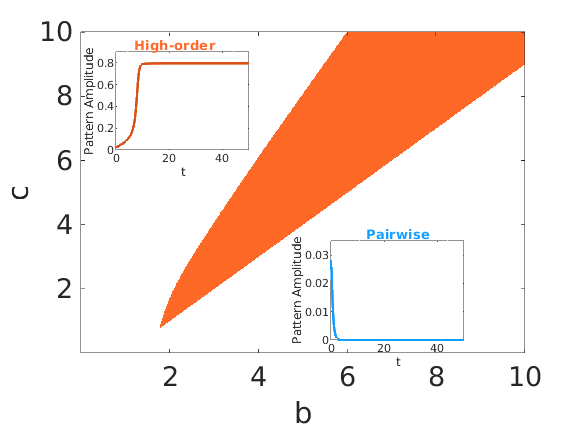}
\includegraphics[width=0.45\linewidth]{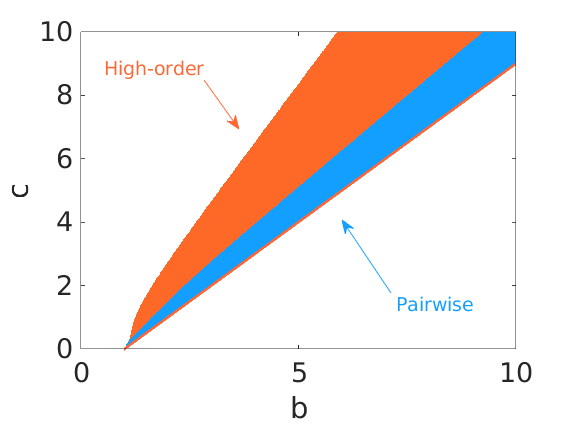}
\caption{Left panel: Brusselator model with $D_u^{(1)}=1$, $D_v^{(1)}=0.5$, $D_u^{(2)}=0.1$, $D_v^{(2)}=2$, $\sigma_1=0.1$ and $\sigma_2=1$. The setting does not allow for Turing patterns when only pairwise interaction are considered, hence to obtain the instability region we need to include also high-order (three-body) interactions (the region of parameters where patterns are obtained is shown in orange). 
Numerical simulations of the system with $b=5.5$ and $c=7$ without (resp. with) high-order terms show that Turing patterns are not obtained (resp. are obtained) as confirmed by the total pattern amplitude \cite{NM2010} in blue (resp. orange) on the lower (resp. upper) inset. Right panel: Brusselator model with $D_u^{(1)}=0.1$, $D_v^{(1)}=0.5$, $D_u^{(2)}=0.01$, $D_u^{(2)}=1$, $\sigma_1=0.2$ and $\sigma_2=1$. The instability region when only pairwise interactions are considered is depicted in blue, while in orange the one obtained when also high-order ones are active; note that the blue region is a subset of the orange one. For both panels, the initial perturbation is $\sim 10^{-2}$ and the hypergraph is that of in Fig. \ref{fig:nat_coupl}.}
\label{fig:mapping}
\end{figure*} 

Based on the above discussion, we thus performed a dedicated numerical analysis of this scenario that allows us to fully exploit the presence of high-order terms. The results are reported in Fig. \ref{fig:mapping} where we show the parameters region associated to a Turing instability in the case of the Brusselator model in Eq.~\eqref{eq:bruss_HG_cubic} for two different sets of values of the the weights $\sigma_i$. On the left panel, the first order diffusion coefficients (i.e., pairwise) do not allow Turing instability (since $(v^{*})^2 D_v^{(1)}>(u^{*})^2 D_u^{(1)}$), on the other hand the presence of high-order diffusion with a suitable choice of the weights $\sigma_i$ allows the formation of patterns. On the right panel, instead, Turing instability arises even in the pairwise setting, but many-body interactions are still beneficial as they yield a larger region of parameters for which patterns can be obtained.

Let us conclude by remarking that, in analogy with the setting studied in the previous section (i.e., regular topologies), it is also possible to find couplings such that high-order interaction hamper the formation of Turing patterns, which would normally arise when the interactions are limited to be pairwise.

\section{Discussion}
\label{sec:disc}

In this work we have formulated a general theory to study the emergence of Turing patterns for dynamical systems where multi-body interactions are taken into account, modeled by using high-order interactions. Our framework goes beyond the one recently proposed~\cite{carletti2020dynamical} that, as shown in the Appendix, can be recovered as a special case of the theory hereby developed. Our framework is inspired by the work of~\cite{gambuzza2021stability} dealing with synchronization in high-order structures and allows to obtain Turing patterns in settings where it would be otherwise impossible, e.g., by restricting to pairwise interactions. At the same time it also permits to suppress the instability which would occur with only pairwise interactions, depending on the desired applications. We have shown that including diffusion terms from high-order interactions may either widen the region of parameters where patterns occur, or, on the contrary, reduce its extension. This can be achieved acting solely on the diffusion coefficients but also using different coupling functions. The detailed analytical study we performed provided us a clear understanding of the role of the model parameters as well as of the high-order topology in the pattern emergence. This further flexibility in the choice of the parameters can shed light on the fine tuning problem \cite{HG2021} and benefit the field of optimal control of patterns \cite{control_hata,control_frasca,control_holme}. In fact, one could choose the appropriate coupling functions and/or diffusion coefficients in order to obtain the desired inhomogeneous state. 

A natural extension of our framework would be to study Turing (wave) instability emerging from a limit cycle: in fact, it has been shown that a similar instability mechanism can occur also when the homogeneous stable state is a limit cycle \cite{chall, entropy}. In many applications where it is important to control the emergence of synchronization, such as neuroscience, our theory could provide to be extremely useful. Moreover, exploring the onset of Turing patterns in the newly developed framework of $M$-directed hypergraphs \cite{gallo2022synchronization} could provide another interesting direction where to focus the attention in future works.

\subsubsection*{Acknowledgements} The authors would like to thank Lorenzo Giambagli and Lucia Valentina Gambuzza for useful discussions and feedback. R.M. is supported by a FRIA-FNRS PhD fellowship, Grant FC 33443, funded by the Walloon region. R.M. and L.G. acknowledge funds from the Erasmus+ program.

\appendix

\section{Formalism mapping}
\setcounter{equation}{0}
\renewcommand{\theequation}{A\arabic{equation}}
\setcounter{figure}{0}
\renewcommand{\thefigure}{A\arabic{figure}}\label{sec:appA}
In this appendix, we show how the formalism introduced in \cite{carletti2020dynamical} to study the formation of Turing patterns on hypergraphs can be recovered from the mathematical framework developed in \cite{gambuzza2021stability} for the analysis of synchronization of chaotic oscillators in simplicial complexes. Note that the latter also applies to the case of hypergraphs, as discussed in \cite{gallo2022synchronization}. 

We start from the dynamical system \eqref{eq:dyn1}. First, we assume the coupling functions $\vec{g}^{(d)}$ to be of the form
\begin{equation}
\begin{array}{lll}
    \vec{g}^{(d)}(\vec{x}_i,\vec{x}_{j_1},\vec{x}_{j_2},\dots,\vec{x}_{j_d}) &=& \varphi(d+1)[\gamma(\vec{x}_{j_1}) - \gamma(\vec{x}_{i}) + \dots + \gamma(\vec{x}_{j_d}) - \gamma(\vec{x}_{i})] \\
    &=& \varphi(d+1)\left[\sum\limits_{n=1}^{d}\gamma(\vec{x}_{j_n}) - d\gamma(\vec{x}_{i})\right]
    \end{array}
    \label{eq:teo_lin_coup}
\end{equation}
where $\varphi:\mathbb{R}\longrightarrow\mathbb{R}$ is a generic function of the number of nodes involved in the high-order interaction, i.e., the size of the hyperedge or of the simplex, while $\gamma:\mathbb{R}^{m}\longrightarrow\mathbb{R}^{m}$ is a generic function encoding the contribution of the node state vectors to the coupling. Note that, given the form in Eq.~\eqref{eq:teo_lin_coup}, all the coupling functions are non-invasive. Eq.~\eqref{eq:dyn1} can be hence rewritten as
\begin{equation}
\begin{array}{lll}
\dot{\vec{x}}_i&=&\vec{f}(\vec{x_i})+\sigma_1 \sum\limits_{j_1=1}^N A_{ij_1}^{(1)}\varphi(2)[\gamma(\vec{x}_{j_1}) - \gamma(\vec{x}_{i})]+ \\ [10pt]
&&+\sigma_2 \sum\limits_{j_1=1}^N \sum\limits_{j_2=1}^N A_{ij_1j_2}^{(2)}\varphi(3)[\gamma(\vec{x}_{j_1}) + \gamma(\vec{x}_{j_2}) - 2\gamma(\vec{x}_{i})] + \dots \\ [10pt]
&&+ \sigma_P \sum\limits_{j_1=1}^N \dots \sum\limits_{j_P=1}^N A_{ij_1j_2\dots j_P}^{(P)}\varphi(P+1)\left[\sum\limits_{n=1}^{P}\gamma(\vec{x}_{j_n}) - P\gamma(\vec{x}_{i})\right]
\end{array}
\label{eq:dyn_teo_lin_coup}
\end{equation}

Noticing that each adjacency tensor $A^{(d)}$ is symmetric with respect to its $d+1$ indices, i.e., $A_{ij_1j_2\dots j_d}^{(d)} = A_{\pi(ij_1j_2\dots j_d)}^{(d)}$, where $\pi$ is a generic permutation of the indices, we can simplify the coupling terms, thus writing Eq.~\eqref{eq:dyn_teo_lin_coup} as
\begin{equation}
\begin{array}{lll}
\dot{\vec{x}}_i&=&\vec{f}(\vec{x_i})+\sigma_1 \sum\limits_{j_1=1}^N A_{ij_1}^{(1)}\varphi(2)[\gamma(\vec{x}_{j_1}) - \gamma(\vec{x}_{i})]+ \\ [10pt]
&&+2\sigma_2 \sum\limits_{j_1=1}^N \varphi(3)[\gamma(\vec{x}_{j_1}) - \gamma(\vec{x}_{i})]\sum\limits_{j_1=1}^N A_{ij_1j_2}^{(2)} + \dots \\ [10pt]
&&+ P\sigma_P \sum\limits_{j_1=1}^N \varphi(P+1)[\gamma(\vec{x}_{j_1}) - \gamma(\vec{x}_{i})] \sum\limits_{j_2=1}^N \dots \sum\limits_{j_P=1}^N A_{ij_1j_2\dots j_P}^{(P)}
\end{array}
\label{eq:dyn_teo_lin_coup_permu}
\end{equation}

By recalling the definition \cite{gambuzza2021stability} of the generalized $d$-degree, $k^{(d)}_{ij}$, of a link $(i,j)$, which represents the number of $d$-order structures the link $(i,j)$ is part of
\begin{equation}
    (d-1)!k^{(d)}_{ij} = \sum\limits_{l_2=1}^N \dots \sum\limits_{l_d=1}^N A_{il_1l_2\dots l_d}^{(d)}
\end{equation}
and by noticing that $k^{(1)}_{ij}=A_{ij}^{(1)}$, we can write the equations governing the dynamics of the system as
\begin{equation}
\begin{array}{lll}
\dot{\vec{x}}_i&=&\vec{f}(\vec{x_i})+\sigma_1 \sum\limits_{j=1}^N k_{ij}^{(1)}\varphi(2)\Delta_j+ 2\sigma_2 \sum\limits_{j=1}^N k_{ij}^{(2)}\varphi(3)\Delta_j + \dots \\&& + P(P-1)!\sigma_P \sum\limits_{j=1}^N k_{ij}^{(P)}\varphi(P+1)\Delta_j
\end{array}
\label{eq:dyn_teo_lin_coup_deg}
\end{equation}
where we have relabeled index $j_1$ as $j$, and where we have defined $\Delta_j = \gamma(\vec{x}_{j}) - \gamma(\vec{x}_{i})$.

By considering the high-order incidence matrix \cite{carletti2020dynamical}, $e_{i\alpha}$, defined as
\begin{align}
e_{i\alpha}&= 
\begin{cases} 1 & v_i \in E_\alpha \\ 0 & \mathrm{otherwise}
\end{cases}
\label{eq:incidence_matrix}
\end{align}
where $E_\alpha$ can represent either a hyperedge or a simplex, we can now write Eq.~\eqref{eq:dyn_teo_lin_coup_deg} as
\begin{equation}
\begin{array}{lll}
\dot{\vec{x}}_i&=&\vec{f}(\vec{x_i})+\sigma_1 \sum\limits_{j=1}^N \sum\limits_{\alpha:|E_\alpha|=2} e_{i\alpha}e_{j\alpha}\varphi(2)\Delta_j+ 2\sigma_2 \sum\limits_{j=1}^N \sum\limits_{\alpha:|E_\alpha|=3} e_{i\alpha}e_{j\alpha}\varphi(3)\Delta_j + \dots \\&& + P!\sigma_P \sum\limits_{j=1}^N \sum\limits_{\alpha:|E_\alpha|=P+1} e_{i\alpha}e_{j\alpha} \varphi(P+1)\Delta_j
\end{array}
\label{eq:dyn_teo_lin_coup_inci}
\end{equation}
In fact, as for each order $d$ of the high-order interactions, the factor $e_{i\alpha}e_{j\alpha}$ is equal to one when both nodes $i$ and $j$ belong to $E_\alpha$, by summing over all $j\neq i$, each summation has exactly $k_{ij}^{(d)}$ non-zero terms. 

Lastly, by assuming the $d$-th coupling strength to be $\sigma_{d} = \varepsilon / d!$, we finally obtain
\begin{equation}
\dot{\vec{x}}_i=\vec{f}(\vec{x_i})+\varepsilon \sum\limits_{j=1}^N \sum\limits_{\alpha} e_{i\alpha}e_{j\alpha}\varphi(|E_\alpha|)[\gamma(\vec{x}_{j}) - \gamma(\vec{x}_{i})]
\label{eq:dyn_teo}
\end{equation}
which recovers the dynamical system analyzed in \cite{carletti2020dynamical} as we set $\varphi(|E_\alpha|) = |E_\alpha|-1$. Hence, the mathematical framework here considered permits to extend the analysis of Turing pattern formation in hypergraphs, as it allows to consider more general coupling functions.

\bibliographystyle{RS}
\bibliography{biblio}

\end{document}